\documentclass{emulateapj}

\newcommand{\Kepler}{{\it Kepler}}
\newcommand{\Spitzer}{{\it Spitzer}}
\newcommand{\hipparcos}{{\it Hippacos}}
\newcommand{\Gaia}{{\it Gaia}}

\newcommand{\be}{\begin{equation}}
\newcommand{\ee}{\end{equation}}

\newcommand{\metallicity}{[M/H]}

\newcommand{\teff}{6199}
\newcommand{\teffe}{50}

\newcommand{\thisnewfephem}{2458271.0740} 
\newcommand{\thisnewfephemerr}{0.0008}

\newcommand{\thisnewdephem}{2458279.709}
\newcommand{\thisnewdephemerr}{0.003}

\newcommand{\mh}{-0.11}
\newcommand{\mhe}{0.08}

\newcommand{\rearth}{R$_\oplus$}

\newcommand{\ldone}{0.311}
\newcommand{\uldone}{0.048}
\newcommand{\ldtwo}{0.31}
\newcommand{\uldtwo}{0.13}

\newcommand{\rprstd}{0.0259}
\newcommand{\urprstd}{0.0015}
\newcommand{\tdurd}{12.71}
\newcommand{\utdurd}{0.26}
\newcommand{\impd}{0.50}
\newcommand{\uimpd}{0.27}

\newcommand{\ttransitd}{2457166.2629}
\newcommand{\uttransitd}{0.0016}

\newcommand{\rprstf}{0.0672}
\newcommand{\urprstf}{0.0013}
\newcommand{\tdurf}{18.998}
\newcommand{\utdurf}{0.051}
\newcommand{\impf}{0.227}
\newcommand{\uimpf}{0.089}

\newcommand{\ttransitf}{2457186.91451}
\newcommand{\uttransitf}{0.00032}

\newcommand{\thisstar}{HIP~41378}

\newcommand{\thisthirdplanet}{HIP~41378~d}

\newcommand{\thisfifthplanet}{HIP~41378~f}

\usepackage{xcolor}
\usepackage{hyperref}
\usepackage{breakurl}
\usepackage{amsmath}
\usepackage{graphicx}
\usepackage{verbatim}
\usepackage{booktabs}

\slugcomment{}

\shorttitle{New Transits of HIP 41378}

\shortauthors{Becker et al.}

\begin{document}


\title{A Discrete Set of Possible Transit Ephemerides for Two Long Period Gas Giants Orbiting HIP 41378}

\email{jcbecker@umich.edu}

\author{Juliette C. Becker$^{1}$, Andrew Vanderburg$^{2,\star}$,  
Joseph E. Rodriguez$^3$,  
Mark Omohundro$^{4}$, 
Fred C. Adams$^{1,5}$, 
Keivan G. Stassun$^{6,7}$, 
Xinyu Yao$^{8}$, 
Joel Hartman$^{9}$, 
Joshua Pepper$^{8}$, 
Gaspar Bakos$^{9}$,
Geert Barentsen$^{10}$, 
Thomas G. Beatty$^{11}$, 
Waqas Bhatti$^{8}$,
Ashley Chontos$^{12}$, 
Andrew Collier Cameron$^{13}$,
Coel Hellier$^{14}$, 
Daniel Huber$^{12}$, 
David James$^{3}$, 
Rudolf B. Kuhn$^{15}$, 
Michael B. Lund$^{6}$, 
Don Pollacco$^{16}$, 
Robert J. Siverd$^{6}$, 
Daniel J. Stevens$^{11,17,18}$, 
Jos\'e Vin\'icius de Miranda Cardoso$^{19}$, 
Richard West$^{16}$, 
}




%

\affil{$^{1}$Astronomy Department, University of Michigan, 1085 S University Avenue, Ann Arbor, MI 48109, USA}
\affil{$^{2}$Department of Astronomy, The University of Texas at Austin, Austin, TX 78712, USA}
\affil{$^3$Harvard-Smithsonian Center for Astrophysics, 60 Garden St, Cambridge, MA 02138, USA}
\affil{$^4$Citizen Scientist}
\affil{$^{5}$Physics Department, University of Michigan, Ann Arbor, MI 48109, USA}
\affil{$^{6}$Department of Physics and Astronomy, Vanderbilt University, 6301 Stevenson Center, Nashville, TN:37235, USA}
\affil{$^{7}$Department of Physics, Fisk University, 1000 17th Avenue North, Nashville, TN:37208, USA}
\affil{$^{8}$Department of Physics, Lehigh University, 16 Memorial Drive East, Bethlehem, PA 18015, USA}
\affil{$^{9}$Department of Astrophysical Sciences, 4 Ivy Lane, Princeton University, Princeton, NJ 08544}
\affil{$^{10}$NASA Ames Research Center, Moffett Blvd, Mountain View, CA 94035, USA}
\affil{$^{11}$Center for Exoplanets and Habitable Worlds, The Pennsylvania State University, 525 Davey Lab, University Park, PA 16802}
\affil{$^{12}$Institute for Astronomy, University of Hawaii, 2680 Woodlawn Drive, Honolulu, Hawaii 96822, USA}
\affil{$^{13}$University of St Andrews, Fife, Scotland}
\affil{$^{14}$Astrophysics Group, Keele University, Staffordshire, UK}
\affil{$^{15}$South African Astronomical Observatory, PO Box 9, Observatory 7935, South Africa}
\affil{$^{16}$Department of Physics, University of Warwick, Coventry, UK}
\affil{$^{17}$Department of Astronomy, The Ohio State University, 140 West 18th Avenue, Columbus, OH 43210, USA}
\affil{$^{18}$Department of Astronomy \& Astrophysics, The Pennsylvania State University, 525 Davey Lab, University Park, PA 16802, USA}
\affil{$^{19}$Center of Electrical Engineering and Computer Science, Federal University of Campina Grande, 58429-900, Brazil}

\affil{$^{\star}$NASA Sagan Fellow} 

\begin{abstract}

In 2015, K2 observations of the bright (V = 8.9, K = 7.7) star HIP 41378 revealed a rich system of at least five transiting exoplanets, ranging in size from super-Earths to gas giants. The 2015 K2 observations only spanned 74.8 days, and the outer three long-period planets in the system were only detected with a single transit, so their orbital periods and transit ephemerides could not be determined at that time. Here, we report on 50.8 days of new K2 observations of HIP 41378 from summer 2018. These data reveal additional transits of the long-period planets HIP 41378 d and HIP 41378 f, yielding a set of discrete possible orbital periods for these two planets. We identify the most probable orbital periods for these two planets using our knowledge of the planets' transit durations, the host star's properties, the system's dynamics, and data from the ground-based HATNet, KELT, and WASP transit surveys. Targeted photometric follow-up during the most probable future transit times will be able to  determine the planets' orbital periods, and will enable future observations with facilities like the \textit{James Webb Space Telescope}. The methods developed herein to determine the most probable orbital periods will be important for long-period planets detected by the \textit{Transiting Exoplanet Survey Satellite}, where similar period ambiguities will frequently arise due to the mission's survey strategy.

\end{abstract}

\keywords{ planets and satellites: detection,  planets and satellites: gaseous planets}

\section{Introduction}

Since the conception of the K2 Mission, one of its major goals has been to detect small transiting planets orbiting nearby bright stars \citep{howell}. The original \Kepler\ mission revealed a diversity of planets with widely varying sizes and equilibrium temperatures. It also showed that small planets, intermediate in size between the Earth and Neptune, are among the most common planets in our Galaxy \citep{Fressin:2013}. Due to \Kepler's narrow and deep survey strategy, most of its discoveries orbit stars that are too distant and faint for detailed follow-up study, so that only limited information can be gleaned about the physical properties of the newly discovered planet population beyond those discernible from the light curves. 

After the original Kepler mission came to an end in 2013, the K2 extended mission \citep{howell} has conducted a series of $\sim$70--80 day observations in different locations along the ecliptic plane. The K2 mission has discovered transiting planets and candidates around bright stars \citep{v15, 2016ApJ...829L...9V, 2017AJ....153..256R, 2017AJ....154..122C, 2017AJ....154..266N, rodriguez18, mayo:2018, yu:2018, Brahm:2018}, which are particularly amenable to follow-up studies, such as precise radial velocities and transit transmission spectroscopy. In particular, the yield from K2 includes six of the ten small planets (with $R_p <3 R_{\oplus}$) with the best prospects for transmission spectroscopy discovered to date \citep{2018arXiv180608368R}. These kinds of follow-up observations could significantly improve our understanding of these planets, yielding information about their interior structure and bulk composition \citep{dressingk93}, how they have been sculpted by processes like photoevaporation \citep{ehrenreich}, how, why and to what extent some planets form aerosols and hazes high in their atmospheres \citep{crossfieldkreidberg, may:2018}, and what kind of molecules constitute their atmospheres \citep{morley}.

One of the most remarkable discoveries from the K2 mission was the HIP 41378 system, a bright (V=8.9, K=7.7) F-type star which was shown to host at least five transiting planets (\citealt{hip41378}, denoted as V16a hereafter). The \thisstar\ planetary system has a rich architecture reminiscent of some of \Kepler's most famous planetary systems \citep{lissauer2014}, but orbits one of the brightest stars found by \Kepler\ to host planets. As such, the HIP 41378 planets remain prime targets for follow-up measurements such as transmission spectroscopy. In particular, HIP 41378 f, with a radius only slightly smaller than Jupiter at 10.2 \rearth, provides a unique opportunity to study in transmission the atmosphere of a temperate gas giant. However, the precise orbital periods of the three outer planets (HIP 41378 d, e, and f) could not be determined from the original Campaign 5 data, since these planets were each only detected with a single transit. Without a precise transit ephemeris, it has not been possible to schedule and carry out transmission spectroscopy observations of these three planets. 

Fortunately, however, K2 had the opportunity to re-observe \thisstar\ during the mission's 18$^{th}$ campaign for an additional 51 days. Here, we analyze these new observations and show that K2 managed to catch two of the long-period planets a second time in transit. We combine previous results from V16a with the new Campaign 18 observations of HIP 41378 to make an updated assessment of the properties of this planetary system. In particular, we identify a set of discrete, precise, possible orbital periods for planets \thisstar\ d and \thisstar\ f, and assess the likelihood that each possible period is the true orbital period. Identifying these possible periods paves the way towards transmission spectroscopy and other follow up observations. Our methods for identifying the most probable orbital periods also demonstrate a strategy for determining a planet's period/ephemeris when multiple transits are observed, but not with sufficient duty cycle for the orbital period to be uniquely determined. This situation has been moderately common in K2, and will be even more important with upcoming observations from the \textit{Transiting Exoplanet Survey Satellite} (TESS). 

This paper is organized as follows. The data from the K2 Campaign 18 are described in Section \ref{sec:k2data}, with a focus on the outer planets \thisstar\ d and \thisstar\ f, which were seen in transit during the observational window. This additional data is used to provide updated constraints on the properties of the \thisstar\ planetary system. 
In Sections 2.2, 2.3 and 2.4, we present observations from the KELT, HATNet and WASP surveys, respectively, which provide additional constraints on the system.
In Section \ref{sec:analysis}, we present results from the analysis of all four data sets, including updated stellar parameters and preliminary period estimations for \thisstar\ d and \thisstar\ f.
In Section \ref{sec:dynamics}, we use a dynamical analysis to place improved constraints on the possible distributions of orbital periods and other orbital elements. The paper concludes in Section \ref{sec:conclude} with a summary of the results and a brief discussion of their implications.

\section{Observations}
 HIP 41378 was observed by \Kepler\ for a total of about 126 days in both Campaigns 5 (C5) and 18 (C18) of the K2 mission. The data from K2 C5 showed evidence of the transits of five planets, three of which transited once each during the original 74-day campaign (V16a). \thisstar\ has also been observed by several ground-based planet-hunting surveys, including the Hungarian-made Automated Telescope (HAT), Kilodegree Extremely Little Telescope (KELT) surveys, and the Wide Angle Search for Planets (WASP).
{We have included as supplemental data to this manuscript the light curves from all five sources that were used in this work.}

\subsection{K2 data}
\label{sec:k2data} 
During C5, HIP 41378 was only observed in long-cadence mode (29.4 minute co-added exposures), but it was observed in short-cadence mode (58.34 second co-added exposures) during Campaign 18 due to the discovery of its planetary system. Analysis of the short-cadence data will likely yield improved parameters for the planets in the system and a detection of asteroseismic oscillations, but we defer this work until the final, pipeline-calibrated data is released by K2 team later in 2018. In this work, we focus on analysis of the long-cadence data to determine precise possible orbital periods for HIP 41378 d and f, with a goal of determining the orbital periods as soon after the last transit of each planet as possible, so that the periods we identify as most likely can be monitored and eventually the true orbital periods will be identified by follow-up work. 


\subsubsection{Campaign 5}

HIP 41378 was observed along with 25,850 other targets by the \Kepler\ space telescope during C5 (2015 April 27 - 2015 July 10) of the K2 mission. Upon downlink of the data, the K2 team processed the data with their photometric pipeline, to produce calibrated pixel files. V16a downloaded the pixel-level data, produced a light curve using the methodology of \citet{vj14}, and then rederived the K2 systematics correction by simultaneously fitting the long-term stellar variability, pointing-related systematics, and transits of the five detected planets following the method of \citet{v15}. We use the highly-precise (38 ppm scatter per 30 minute exposure) light curve produced from the simultaneous fit by V16a for our analysis. The C5 light curve is plotted in the second panel of Figure \ref{fig:fulldata}.

\subsubsection{Campaign 18}

HIP 41378 was observed along with 20,419 other targets  by the \Kepler\ space telescope during C18 (2018 May 12 -- 2018 July 02) of the K2 mission. After the data was downlinked from the spacecraft, the raw cadence files were immediately uploaded to the Mikulski Archive for Space Telescopes (MAST), before pixel-level calibration had been performed by the K2 team. In the interest of time, we used the raw, un-calibrated cadence files to produce a quick-look light curve of HIP 41378. We downloaded the cadence files from the MAST and used the \textit{kadenza} software tool \citep{kadenza} to produce a pseudo-target pixel file containing the long-cadence \Kepler\ images from the postage stamp region around HIP 41378. We then used the procedure of \citet{vj14} to process the K2 pixel data into a systematics-corrected light curve. We manually identified and excluded cadences from our systematics correction when \Kepler\ was undergoing a reaction wheel desaturation event. We also manually excluded a continuous stretch of seven hours of data around time BJD - 2454833 = 3431.85 when \Kepler\ experienced a pointing anomaly. The systematics corrected light curve showed transits of four of the five known HIP 41378 planets: HIP 41378 b, c, d, and f. After performing a first-pass systematics correction with the \citet{vj14} method, we re-derived the systematics correction and re-processed the light curve following the method of \citet{v15} to simultaneously fit for the long-term variability, pointing-related systematics, and the transits of the four planets seen in C18. The photometric precision of the light curve is about 40\% worse (51 ppm scatter per 30 minute exposure) than the light curve from C5 as a result of using the un-calibrated pixel data. 

\begin{figure*} 
   \centering
   \includegraphics[width=7.0in]{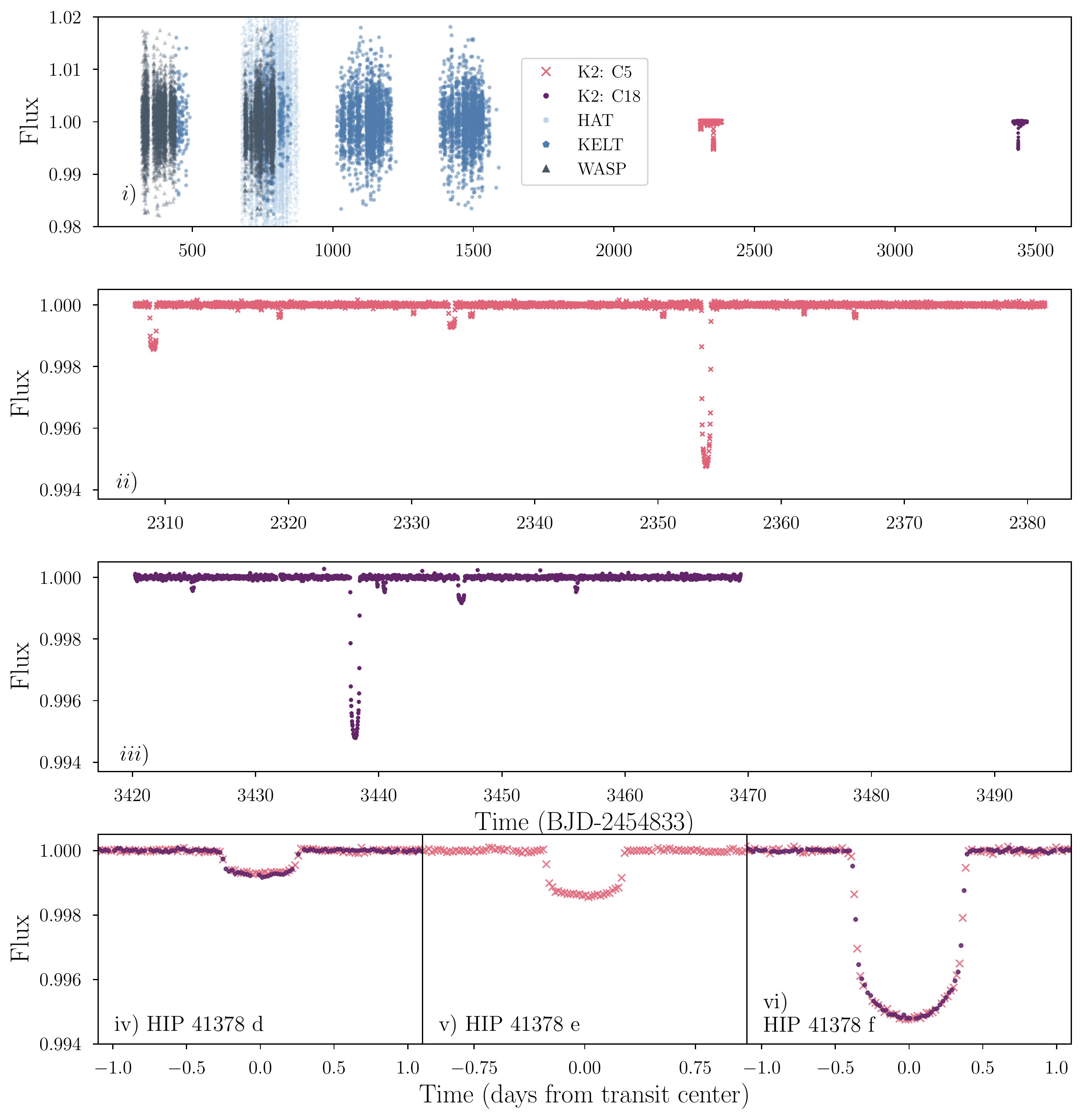} 
   \caption{\thisstar\ was observed in C5 and C18 of the K2 mission, which were separated by a little over 1000 days. (top panel) The K2 data, plotted by time of acquisition, plotted alongside the KELT, HAT, and WASP observations of the same star. All values have been converted to relative flux. (middle two panels) The K2 campaigns, expanded. In both campaigns, many transits of \thisstar\ b and c are observed, while single transits of \thisstar\ d and f are observed in each campaign. \thisstar\ e transits only once in the first campaign (C5). (bottom panels) The phase-folded light curves for planets \thisstar\ d, e and f, each of which transited only once during the K2 C5. Although we do not know the true orbital period of any of these three planets, we plot the results of a Levenberg-Marquardt fit of the transit model to the transit center for each event and center the plot on each fitted center. Planets d and f have data from both C5 and C18, while e has no data from C18 since it did not transit during this newer campaign. }
   \label{fig:fulldata}
\end{figure*}

{After submission of this paper, the K2 team released their pipeline-processed target pixel files from Campaign 18. We downloaded these newly processed data and analyzed them in an identical manner to the Campaign 5 data, following \citet{vj14} and \citet{v15} to extract light curves, produce a first-pass systematics correction, and then fit simultaneously for the systematics correction, transit model, and low-frequency variability. The resulting light curve has photometric precision nearly identical to that of the Campaign 5 light curve (38 ppm per 30 minute exposure). This re-processed light curve is plotted in the second panel of Figure \ref{fig:fulldata}, and we use this updated light curve in the rest of our analysis.}

\subsection{KELT data}
\label{sec:kelt}
The Kilodegree Extremely Little Telescope (KELT) survey \citep{2007PASP..119..923P, pepper:2012} is a ground-based transit survey consisting of two 4.2cm-aperture, wide-field (26 by 26 degrees) automated telescopes (KELT-North is located in Arizona, and KELT-South is located in South Africa). 
KELT's primary goal is the detection of transiting Jupiter-sized planets, and it has had much success finding these planets \citep{2016AJ....152..136Z,2017AJ....153..178S,  2017AJ....154..194L, 2017Natur.546..514G, 2017AJ....153..215P, 2018AJ....155...35S}.
The candidate selection process is described in \citep{2012ApJ...761..123S} and \citep{2016MNRAS.459.4281K}; 
KELT is primarily searching for planets with V-band magnitudes between 8 and 10, and HIP 41378 was observed by both KELT telescopes for several years before Kepler's K2 mission began.
Roughly 4700 observations of \thisstar\ were taken by KELT-North and KELT-South between March 2, 2010, and May 10, 2013.
The reduction pipeline and candidate selection process is described in \citet{2012ApJ...761..123S} and \citet{2016MNRAS.459.4281K}.
The full KELT data set for \thisstar\ is plotted in Figure \ref{fig:fulldata}. 



%
%
%
%
%

%
%
%
%
%

\subsection{HATNet data}
\thisstar\ was a target observed by HATNet \citep{2004PASP..116..266B} between November 2, 2010 and June 3, 2011. 
HATNet is a telescope network which consists of six 11-cm, wide-field (10.6 degrees by 10.6 degrees field of view) aperture lenses on six different fully-automated telescope mounts, four of which are in Arizona and two of which are in Hawaii \citep{2004PASP..116..266B}.
The HATNet observations for \thisstar\ were reduced as in \citet{2010ApJ...710.1724B}, using aperture photometry routines from image processing software FITSH \citep{2012MNRAS.421.1825P}. 
The resultant light curves were outlier-clipped, smoothed, and detrended using the TFA \citep{2005MNRAS.356..557K}. 
Due to the brightness of HIP 41378 the innermost pixels containing the center of the point spread function (PSF) are saturated in the HATNet observations of this star. Because the Apogee U16m 4K$\times$4K CCDs used by HATNet incorporate anti-blooming gates, electrons exceeding the full pixel well are suppressed rather spilling over into neighboring pixels. To account for this, aperture photometry is performed only on the unsaturated pixels and then corrected for the flux not counted in the saturated pixels using the estimated PSF. This leads to bright/saturated stars having lower photometric precision than somewhat fainter/unsaturated stars. For HIP 41378 the HATNet observations have a RMS scatter of 14\,mmag, compared to $\sim 4$\,mmag for the brightest unsaturated stars in the same images.

The full HAT data set for \thisstar\ is also plotted in Figure \ref{fig:fulldata}, and consists of 12903 observations.

\subsection{WASP data}
\thisstar\ was also target observed by the WASP survey \citep{Pollacco:2006} between Nov 20, 2009 and March 3, 2011.
WASP consists of two robotic telescope arrays with eight Canon lenses (each with a field of view of 7.8 degrees by 7.8 degrees). The arrays are located in South Africa and the Canary Islands. 
Data taken post January 2009 benefit from improved red noise reduction \citep{Barros:2011,Faedi:2011}.
WASP data are detrended using SysRem \citep{Tamuz:2005} and TFA.

The WASP light curve for \thisstar\ was further processed: all points with error bars greater than 2\% were excluded, as were points with flux values smaller than 10\% the median flux value. The remaining points include are expected to be good quality, and are plotted in Figure \ref{fig:fulldata}.

\section{Analysis}
\label{sec:analysis}
\subsection{Updated Stellar Parameters}
Our analysis to determine the most likely orbital periods of the long-period HIP 41378 planets depends directly on the adopted stellar parameters, especially the stellar density. V16a used observations from the TRES spectrograph on the 1.5m telescope at Mt. Hopkins, AZ to derive spectroscopic properties (including stellar metallicity, temperature, and V band magnitude) using the Stellar Parameter Classification \citep[SPC, ][]{buchhave, buchhave14} method. These values are reproduced in Table \ref{tab:1}. 
V16a determined fundamental stellar parameters (like the stellar mass and radius) using a parallax from the \hipparcos\ mission, but since the publication of their work, the \textit{Gaia} mission has released significantly more precise parallax measurements \citep{gaia1, gaia2}. 
Using the updated parallax from \textit{Gaia} Data Release 2 (reported in Table \ref{tab:1}), and the previous measured values of stellar metallicity, temperature, and V band magnitude, we used an online interface\footnote{\url{http://stev.oapd.inaf.it/cgi-bin/param_1.3}} to interpolate onto Padova stellar evolution tracks \citep{dasilva}. 
The resulting, updated values of stellar mass, stellar radius, and surface gravity are presented in Table \ref{tab:1}; thanks to the new, precise \Gaia\ parallax, we are able to determine the stellar density, an important quantity for our dynamical analysis, with much higher precision than V16a.

We obtained an independent, empirical measure of the stellar radius via the method described by \citet{Stassun_etal:2018}. Briefly, we performed a fit to the full broadband spectral energy distribution (SED) using a Kurucz model atmosphere with the same stellar $T_{\rm eff}$, $\log g$, and [Fe/H], and their uncertainties, reported in Table~\ref{tab:1}. The free parameters of the SED fit are then only the extinction ($A_V$) and an overall flux normalization. The bolometric flux at Earth ($F_{\rm bol}$) is then obtained simply by direct summation of the (non-reddened) SED model. Finally, the stellar radius then follows from the $F_{\rm bol}$ and $T_{\rm eff}$ via the Stefan-Boltzmann relation. 

We constructed the observed stellar SED using broadband fluxes spanning 0.2--20~$\mu$m from {\it GALEX\/} NUV, {\it Tycho-2\/} $B_T V_T$, APASS $BVgri$, 2MASS $JHK_S$, and {\it WISE\/} 1--4. We limited the maximum permitted $A_V$ to be that of the full line-of-sight extinction from the Galactic dust maps of \citet{Schlegel:}. The resulting best fit SED model, with a reduced $\chi^2$ of 2.9, gives $A_V = 0.01 \pm 0.01$ and $F_{\rm bol} = 6.91 \pm 0.16 \times 10^{-9}$ erg s$^{-1}$ cm$^{-2}$. Adopting the {\it Gaia\/} DR2 parallax, but adjusted by +0.08~mas to account for the systematic offset found by \citet{Stassun_Torres:2018}, we obtain $R_\star = 1.35 \pm 0.02$~R$_\odot$, consistent with the radius from our analysis using the Padova isochrones. For the rest of our analysis, we adopt the parameters from the isochrone analysis (reported in Table \ref{tab:1}). 


\subsection{Measured center of transit times for new transits of \thisstar\ d and \thisstar\ f}

We measured the mid-time of the two newly detected transits of \thisstar\ d and \thisstar\ f in the C18 data in order to precisely determine the time between the transit observed in C5 and the newly detected transit from C18. 
We determined mid-transit times for the new transits of HIP 41378 d and f by fitting the C18 light curve using a \citet{mandelagol} transit model using a Levenberg-Marquardt least-squares minimization algorithm. Because the un-calibrated C18 data is of somewhat lower quality than the fully processed C5 data, we only performed the Levenberg-Marquardt minimization over one free parameter, the mid-transit time, while fixing the transit shape parameters ($R_{p}^2/R_{*}^2$, transit duration, impact parameter, and quadratic limb darkening parameters) to the values reported by V16a. 

The resulting mid-transit times from our one-parameter optimizations of the C18 light curve and the mid-transit times determined by V16a from C5 are given in Table \ref{tab:1}. We defer further refinement of the system parameters until the fully processed C18 light curve is available. 

\begin{deluxetable*}{lcccr}
\tablecaption{Relevant System Parameters for \thisstar \label{bigtable}}
\tablewidth{0pt}
\tablehead{
  \colhead{Parameter} & 
  \colhead{Value}     &
  \colhead{}   &
  \colhead{} 
}
\startdata

\emph{Stellar Parameters} & &  \\
Right Ascension & 8:26:27.85 & &   \\
Declination & +10:04:49.35 & &   \\
Age~[Gyr] & 4.279 $\pm$ 0.931   & &   \\
Parallax~[mas]& 9.379940 $\pm$ 0.059038\\
V magnitude & 8.93 \\ 
$M_\star$~[$M_\odot$] & 1.140 $\pm$ 0.033 \\
$R_\star$~[$R_\odot$] & 1.35 $\pm$ 0.03  \\
(B-V) & 0.599 $\pm$ 0.012 mag \\
Limb darkening $q_1$~ & \ldone   $\pm \uldone$ \\
Limb darkening $q_2$~ & \ldtwo   $\pm \uldtwo$  \\
$\log g_\star$~[cgs] & 4.20 $\pm$ 0.03 \\
Metallicity \metallicity & \mh  $\pm\mhe$ \\
$T_{\rm eff}$ [K] & \teff  $\pm \teffe$\\
 & & \\

\emph{\thisthirdplanet} & & & \\
Radius Ratio, $(R_P/R_\star)$ & \rprstd  $\pm \urprstd$  \\
Transit Impact Parameter, $b$ & \impd $\pm \uimpd$  \\
Time of Transit (C5) $t_{t,5}$~[BJD] & \ttransitd  $\pm$ \uttransitd \\ 
Time of Transit (C18) $t_{t,18}$~[BJD] & \thisnewdephem  $\pm$ \thisnewdephemerr  \\ 
Transit Duration  $D$~[hours] & \tdurd  $\pm$ \utdurd \\ 
& & \\

\emph{\thisfifthplanet} & & & \\
Radius Ratio, $(R_P/R_\star)$ & \rprstf  $\pm \urprstf$  \\
Transit Impact Parameter, $b$ & \impf $\pm \uimpf$  \\
Time of Transit (C5) $t_{t,5}$~[BJD] & \ttransitf  $\pm$ \uttransitf\\ 
Time of Transit (C18) $t_{t,18}$~[BJD] & \thisnewfephem $\pm$ \thisnewfephemerr  \\ 

Transit Duration $D$ [hours] & \tdurf  $\pm$  \utdurf \\ 
 & & \\
\enddata
\tablecomments{Parameters for host star \thisstar\ and considered planets \thisstar-d and \thisstar-f. Planetary data in this table comes from analysis in \citet{hip41378}, and stellar properties ($M_\star$, $R_\star$, stellar age, B-V magnitude, and $\log g_\star$)  have been updated with values using the Gaia parallax (9.37993950 mas $\pm$ 0.059037831). The time of transits are reported for both campaign 5 (C5) and campaign 18 (C18). {For brevity, we do not reproduce the entire table of planetary parameters from \citet{hip41378}, but all parameters not listed in this table and yet used in our present analysis were drawn from the distributions reported in that work. }}
\label{tab:1}
\end{deluxetable*}

\subsection{Period Constraints from the K2 baseline and transit likelihood}
\label{baseline_out_of_transit}
Planets \thisstar\ b and \thisstar\ c have well-determined periods, as they transited multiple times in the C5 Kepler data. We also detect these two planets in our C18 dataset (three transits of \thisstar\ b and one transit of \thisstar\ c), but we defer a full analysis of these two planets {to} future work. Here, we focus our analysis on determining the orbital periods of \thisstar\ d and \thisstar\ f, which is a particularly promising target for transmission spectroscopy of a cold Jupiter, if its times of future transit could be determined. 

In V16a, when we only had detected a single transit each of \thisstar\ d, e, and f, we were able to place broad constraints on their orbital periods. In the C18 data, \thisstar\ d and f each transit one more time each, yielding a discrete spectrum of possible orbital periods. Here, we combine broad constraints on the orbital periods and the discrete possible periods based on the times of the two detected transits to determine the most likely orbital periods for \thisstar\ d and f. 

\begin{figure} 
   \centering
   \includegraphics[width=3.4in]{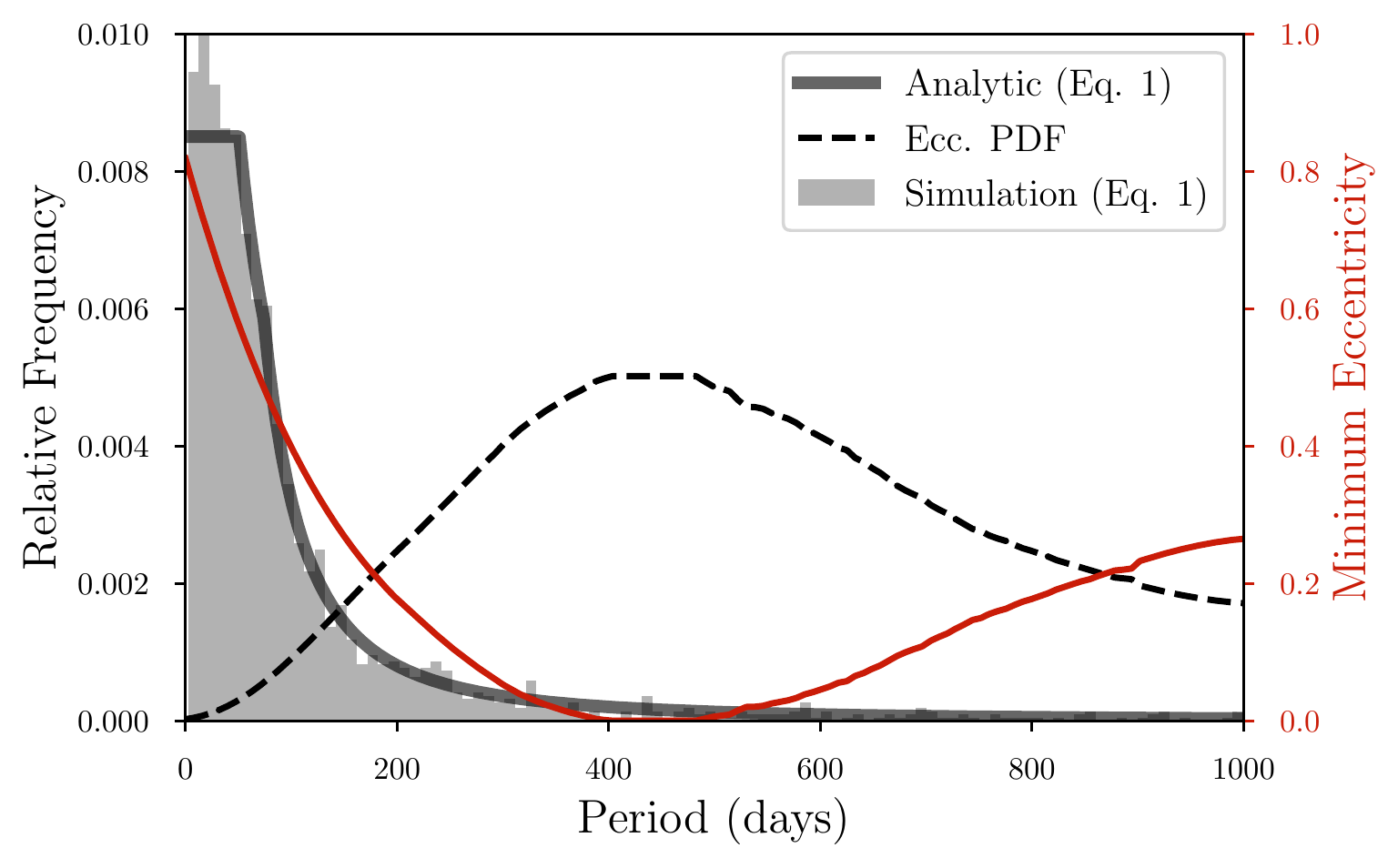} 
   \caption{We show the normalized probability density function for the chance of observing \emph{at least} two transits, with at least one in each of two K2 campaigns, for some planet. Plotted here is a comparison of the analytically derived Equation \ref{transitlhood} (solid line) with a numerically computed Monte Carlo simulation of observability probability by orbital period (histogram).  The distributions plotted here were computed for planet \thisstar\ f, for two K2 campaigns with baselines $B_{1} = 74$ days and $B_{2} = 51$ days, separated by some unspecified length of time. This generalized form describes the chance that only two transits of a planet would be seen over two K2 campaigns, and the good agreement between the simulation and analytic form suggests that Eq. \ref{transitlhood} can be adopted as the baseline prior in cases where a planet with an unknown period is observed over multiple K2 campaigns. Also plotted in comparison is the minimum value of eccentricity (in red) at each orbital period that allows recreation of the observed duration, and (black dotted line) the probability distribution for each eccentricity, as derived from the cumulative density function of the Kipping beta distribution eccentricity prior (where all values larger than the minimum eccentricity at each period are considered able to reproduce the true transit duration).  }
   \label{fig:c18base}
\end{figure}

Our analysis to place broad constraints on the orbital periods of these planets closely follows that of V16a, with a handful of differences. In particular, V16a imposed a transit  prior on the calculated orbital periods to account for the fact that planets \thisstar\ d, e, and f transited once during the 74-day baseline of C5 (see Equation 3 of V16a). Now that we have re-detected planets d and f in C18, we can update our prior on the orbital periods based on these new observations. As such, we impose a prior on both planets orbital periods which account for the fact that both HIP 41378 d and f were detected during both C5 and C18:
\begin{equation}
\begin{aligned}
\mathcal{P}_i(P_{i}, D_{i}, B_{5}, B_{18}) = 
\qquad \qquad \qquad \qquad \qquad \qquad \\
\begin{cases}
1 \ \ \ {\rm if}\ P_{i} - D_{i} < B_{5} \\
(B_{5} + D_{i}) / P_{i} \ \ \ {\rm if}\ B_{18} < P_{i} - D_{i} < B_{5}\ \\
(B_{5} + D_{i})(B_{18} + D_{i}) / P_{i}^{2} \ \ \ {\rm else},\ \\
\end{cases}
\end{aligned}
\label{transitlhood}
\end{equation}
\noindent where $\mathcal{P}_i$ is the chance of seeing planet $i$, $B_k$ the time baseline of the observations for campaign $k$, $D_i$ the planet's transit duration, and $P_i$ the orbital period of the planet in question. {This expression can be generalized to apply to any number of distinct campaigns.}
In Figure \ref{fig:c18base}, we show the comparison between this analytic prior and a Monte Carlo simulation of transit probabilities for 20000 randomly chosen periods on the interval $(0,1000]$ days, with random centers of transit times on the same interval, and with randomly selected baseline separations (defined as the time between the last data point of the first campaign with baseline $B_{1}$ and the first data point of the second campaign with baseline $B_{2}$) on the range $[0,3000)$ days. The true separation between the end of C5 and the start of C18 was 1037.13 days, but Equation \ref{transitlhood} describes the general transit probability for a planet with a given period which transits only twice in the K2 data: once in one campaign, and once in another. 
This prior (Eq. \ref{transitlhood}) describes generally the relative chances of planets with various orbital periods transiting over two observing campaigns. This general result can be applied to targets with unknown periods seen over multiple K2 campaigns, and will also be applicable to similar planets observed in two separated baselines by TESS. 
In this work, we will apply this probability as a prior in our calculation of the likelihood of each possible period for \thisstar\ d and f, as shorter period planets are more likely to transit over the observed baselines. 

As done in V16a, we can find lower limits on the periods of \thisstar\ d and \thisstar\ f from then length of out-of-transit observations taken on either side of each event (see the middle two panels of Figure \ref{fig:fulldata} to see visually the out-of-transit baseline on each side of each transit event). In C5, the data was acquired at times (BJD-2454833) between 2307.55 and 2381.41, for an official baseline of 73.86 days\footnote{The total time baseline of C5 was 74.8 days, but V16a clipped out the first day of data while \Kepler\ was thermally settling into the campaign.}. For C18, the data was acquired at times (BJD-2454833) between 3418.56 and 3469.40 days, for a total baseline of 50.86 days. We assumed that for a `detection' to occur, any part of the planet's ingress or egress must have occurred during the K2 observations. 
Using these times, as well as the center of transit times and duration for each transit event, we can compute the minimum periods which are allowed given the single detection in each campaign as follows:
\begin{equation}
\begin{aligned}
P_{min} = \rm{max}(t_{c} + D/2 - B_{i},\ B_{f} - t_{c} + D/2) 
\label{eq:minper}
\end{aligned}
\end{equation}
where $B_{i}$ denotes the beginning time of the campaign, $B_{f}$ denotes the end time of the campaign, $t_{c}$ denotes the measured center time of transit for the observed transit event, and $D$ denotes the planetary transit duration.
The C5 center for \thisstar\ d (with transit duration 12.71 hours) was 2457166.2629 BJD, and the C18 center was \thisnewdephem\ BJD.
Similarly, the C5 center for \thisstar\ f (with transit duration 18.998 hours) was 2457186.91451 BJD, and the C18 center was \thisnewfephem\ BJD.

From this, we compute a minimum period of 48.1 days for \thisstar\ d and a minimum period of 46.4 for \thisstar\ f. Both of these limits come from the C5 data, which had a longer observational baseline. From these limits, we can exclude any periods for these planets less than these values: if the true periods were smaller than these values, we should have seen evidence of a second transit in the C5 data.


Finally, the fact that \thisstar\ d and \thisstar\ f transit twice allows us to define their orbital periods as
\begin{equation}
    P_{j} = \frac{t_{j, 18} - t_{j, 5}}{i},
    \label{eq:simpleperiods}
\end{equation}
when $j$ denotes the planet and $i$ is some positive integer. This equation, combined with the lower limits previously derived, provides a discrete set of possible orbital periods for each planet. For example, there are 23 possible orbital periods for \thisstar\ f, ranging between $\sim$47.1 days (the lowest possible value that exceeds our lower limit on the period) and $\sim$1084 days (if no intermediate transits occurred in between the two we observed). The possible periods for each planet are given in the first column of Table \ref{tab:3} and Table \ref{tab:5} respectively.

\section{Refining Planetary Orbital Periods}
\label{sec:dynamics} 
\subsection{Excluding orbital periods using all HAT/KELT/WASP data}
In Equation \ref{eq:simpleperiods}, we gave an expression for the possible orbital periods for each planet, based on the time at which the two observed events occur for each planet. 
The HAT/KELT/WASP data subtend a significant observational baseline, and the transit event of \thisstar\ f is relatively deep (0.5\%). As such, these ground-based surveys should be able to detect the transit, if the transit happened to occur while the surveys were observing the star.
There are a large number of possible orbital periods {for \thisstar\ f}, so we evaluate the KELT/HAT/WASP evidence for each {possible} orbital period by computing the likelihood ratio for two models: a flat line and the transit model. 
{We chose our outlier rejection threshold so that on average, we reject only one `good' data point from each photometric dataset. We found the significance level that corresponds to the single most extreme data point in each data set (where the KELT data had 4709 unique points, HAT had 12903 points, and WASP had 6732 points), and then utilized sigma-clipping to remove outliers. The threshold we use is 3.7/3.9/3.8$\sigma$ (for KELT/HAT/WASP) away from the median flux level of each survey, which was computed from the number of data points for each survey. }
Then, we phase-folded the light curve at each possible orbital period, and computed the following likelihood ratio:
\begin{equation}    \label{eq:lhoodrat}
\mathcal{L} = \dfrac{\exp{[-\sum_{i=0}^{i=N}(0.5 (f(t_{i}) - m(t_{i}))^{2} \sigma_{i}^{-2}})]}{\exp{[-\sum_{i=0}^{i=N}(0.5 (f(t_{i}) - \bar{f})^{2} \sigma_{i}^{-2}})]}
\end{equation}
where $f(t_{i})$ denotes the KELT, HAT, or WASP flux at each exposure time $t_i$, $m(t_{i})$ denotes the transit model (using the best-fit parameters from Table \ref{tab:1} and the transit model from \citealt{mandelagol}) at some exposure time, $\bar{f}$ denotes the weighted mean KELT/HAT/WASP flux \citep[a flat line model set to be the weighted mean of the out-of-transit baseline, which is expected to be the best model if no transit occurs;][]{kovacs:2002}, and $\sigma_i$ denotes the error on the flux measurements, which is set to be $\sigma_i = k \sigma_{\rm{obs,i}}$, where the measured errors $\sigma_{\rm{obs}}$ for all data points of the each data set have independently been scaled by a factor of $k$, so that the reduced $\chi^{2}$ for the flat line model for each full data-set is roughly 1. The errors on the KELT, HAT, and WASP data-sets are derived and scaled independently. 

\begin{figure*} 
   \centering
   \includegraphics[width=7.0in]{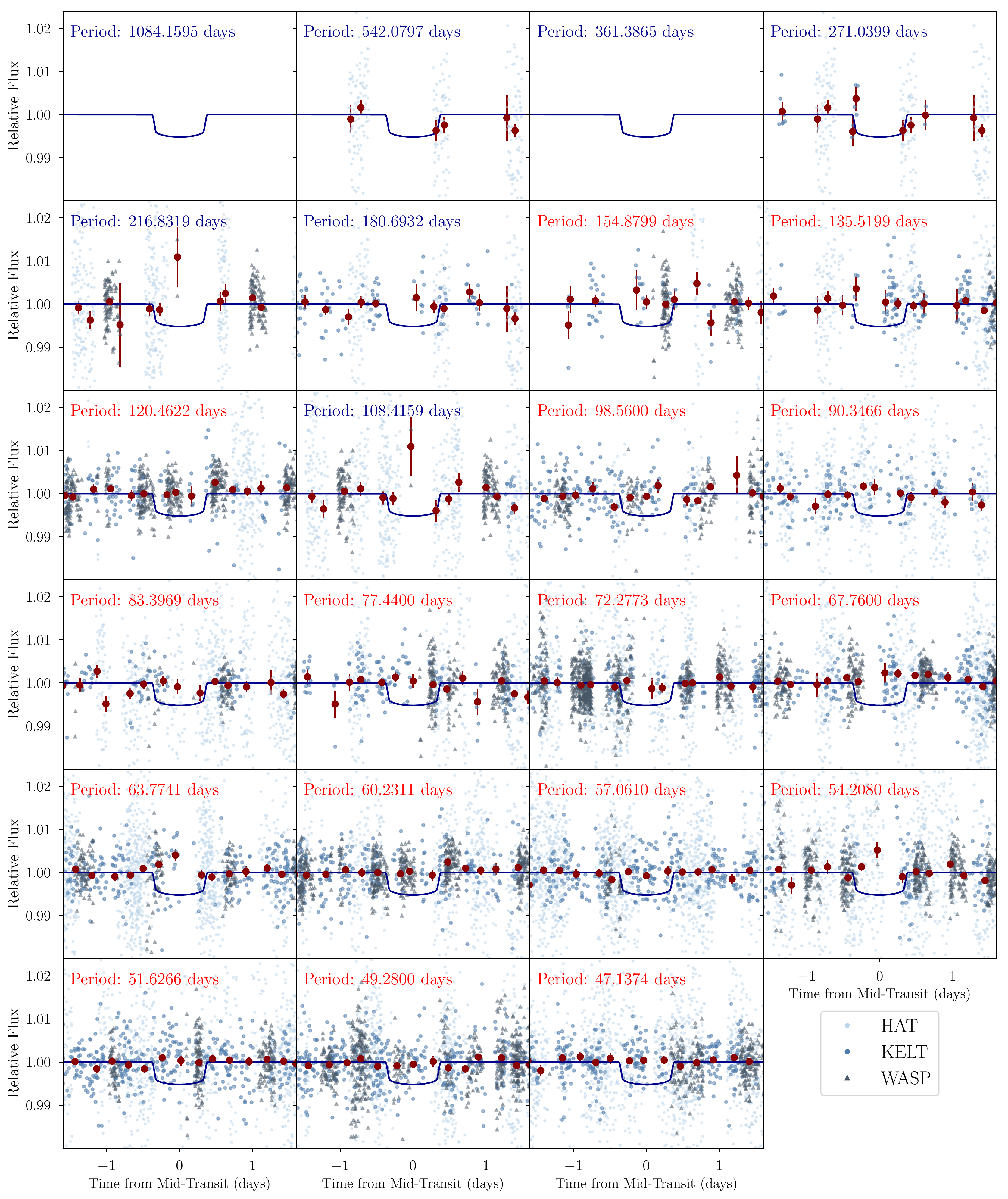}
   \caption{For each possible orbital period of \thisstar\ f derived from Equation \ref{eq:simpleperiods}, we phase-fold the KELT, HAT, and WASP data and fit two models around where the transit would be expected to be: the transit model, plotted in blue, which we would expect to see if the tested orbital period were the true orbital period of the planet; and the best-fit flat model with no transit (not shown), which we expect to see if that orbital period is incorrect. We then compute the likelihood ratio between the two cases to determine whether a flat model is preferable to the transit model. In red are the orbital periods with likelihood ratios of 10000 or more, where the flat line model is heavily preferred. In blue are the orbital periods for which a determination between the two models cannot be made. Blue points are true data points from the KELT, HAT, and WASP surveys, and red points are the weighted mean for each bin, with errors equal to the weighted error on the mean. We note that for a single orbital period - $\sim$ 180 days - the likelihood ratio was roughly 3000, a marginal case which our strict criterion of rejection ($\mathcal{L} > 10000$) does not reject. }
   \label{fig:keltfolded}
\end{figure*}

The likelihood ratio defined in Eq. \ref{eq:lhoodrat} is the ratio of the likelihood of the flat line model to the likelihood of the transit model. In Figure \ref{fig:keltfolded}, we show the phase folded HAT/KELT/WASP light curve at each orbital period. 
For each orbital period, we compute the likelihood ratio using the entire HAT/KELT/WASP combined, phase-folded data-set. The result is one likelihood ratio $\mathcal{L}$ for each orbital period, describing the relative likelihood of the two models. For likelihood ratios $\mathcal{L}$ greater than $10^{4}$, where the flat line model is highly preferred to the transit model, we consider the corresponding orbital period ruled out. 

In each panel of Figure \ref{fig:keltfolded}, we label the orbital period depicted and color the text corresponding to the computed value of the likelihood ratio $\mathcal{L}$: red text indicates that an orbital period can be considered ruled out, and blue text indicates that a particular period cannot be ruled out. For likelihood ratios less than $10^{4}$, we consider the evidence too weak to discriminate between the models. 
Our choice of $10^{4}$ as the significant likelihood ratio was purposefully conservative, decreasing the probability of incorrectly rejecting a particular orbital period.
For the orbital periods we could rule out based on this test, we set the probabilities to be unlikely (defined as $<0.1\%$) in Table \ref{tab:5}. The analysis of the combined KELT, HAT, and WASP data allowed us to eliminate 16 of the 23 possible orbital periods for \thisstar\ f to this significance level. We note that analysis of any of the individual data sets alone could not rule out all 16 orbital periods: the full result of this method was achieved by combining the three ground-based data sets. 

We also performed the same algorithm described above on the predicted orbital periods for \thisstar\ d, but no periods could be excluded (as expected for a transit event of the much smaller measured depth of \thisstar\ d). 
{The depth of \thisstar\ d is $\sim$1 mmag, and its duration $\sim$ 12.7 hours. For the three ground-based surveys considered, typical scaled photometric uncertainties were reported to be 0.5\%, 0.3\%, and 1\% for KELT, WASP, and HAT, respectively, which we roughly corresponds to a median precision of 5 mmag. For a best-case 100 points in transit at this precision, this corresponds to a SNR of $\sqrt{100}(1\rm{\ mmag} / 5\ mmag) = 2$. Once the orbital period for \thisstar d is uniquely determined, or more ground-based data points become available, this signal may be detectable in the HAT/KELT/WASP data. }

Additionally, we conducted a pre-recovery search in the KELT data using the fixed-duration and fixed $T_C $ BLS method, as implemented in the VARTOOLS package \citep{vartools}, with an approach described in \citet{yao:2018}. 
Using the transit time and transit duration for \thisstar\ f as determined from V16a, we searched 300,000 trials evenly spaced in frequency from 150 to 450 days. The strongest peak in the BLS output corresponds to 328.59 days, with a signal-to-pink noise statistic of 2.4, which is below the 10\% confidence level for a transit with the corresponding depth and duration, as determined in \citet{yao:2018}. As such, this is not considered a plausible detection.

\subsection{Dynamically Feasible Periods}

From the K2 data, we have well-measured values of the transit duration, transit impact parameter, center time of transit, and $(R_P/R_\star)$ for \thisstar\ d and \thisstar\ f. These known values are presented in Table \ref{tab:1}.
Using these values and priors on the unmeasured quantities (orbital eccentricity and longitude of perihelion), we can estimate the orbital period that corresponds to a transit duration of a given value using \citep[e.g.,][]{seager2003, duration_ford}:
\begin{multline} \label{eq:duration}
D_{i} = \frac{P_{i}}{\pi} \arcsin{}\left[\left(\frac{G (M_{*} + m_{p,i}) P_{i}^{2}}{4 \pi^{2}}\right)^{-1/3} \times \right. \\
\left. \sqrt{(R_{P,i} + R_{*})^2 - (b_{i}^2 \times R_{*}^2)} \right] \frac{\sqrt{1-e_{i}^2}}{1+e_{i} \cos{\varpi_{i}}}
\end{multline}
where we define $D_{i}$ is the transit duration of the $i$th planet (from first to fourth contact) and $R_{P,i}$ is the planetary radius, $P_{i}$ its orbital period (the quantity for which we would like to solve),  $m_{p,i}$ the mass of the $i$th planet, $e_i$ the orbital eccentricity, $\varpi_i$ the longitude of periastron, $b_i$ the impact parameter, $M_{*}$ the stellar mass, $R_{*}$ the stellar radius, and finally $G$ the gravitational constant. 

In a method reminiscent of the analysis in V16a, we generated 10000 feasible orbital periods by solving the above equation with draws from the following probability distributions:
$t_{d,i}$, $R_{P,i}$, and $b_i$ were drawn from the posteriors given in Table \ref{tab:1}; $M_{*}$ and $R_{*}$ were drawn from the Gaia posterior probability distributions (see also Table \ref{tab:1}); eccentricity $e$ was drawn from a beta distribution with shape parameters $\alpha = 0.867$ and $\beta = 3.03$ \citep[][]{kipping_prior1, kipping_prior2, kipping16}; $\varpi_i$ was chosen using Equation 19 of \citet{kipping16}; $m_{p,i}$ was drawn from the \citet{weissmarcy} mass-radius relation for planets with $R_{p} <1.5 R_{\oplus}$, drawn from the \citet{angie} relation for planets with $4 R_{\oplus} > R_{p} > 1.5 R_{\oplus}$, and the  mean planetary density is drawn from a normal distribution centered at $\rho = 1.3 \pm 0.5$ g / cm$^{3}$ for planets larger than 4\rearth. 
With only $P_{i}$ left as a free parameter in the equation, we solve Eq. \ref{eq:duration} numerically for each set of draws. 
The resultant series of 10000 orbital periods all geometrically produce the observed transit durations, and can as such be considered plausible. 

\begin{figure*} [t!]
   \centering
   \includegraphics[width=6.0in]{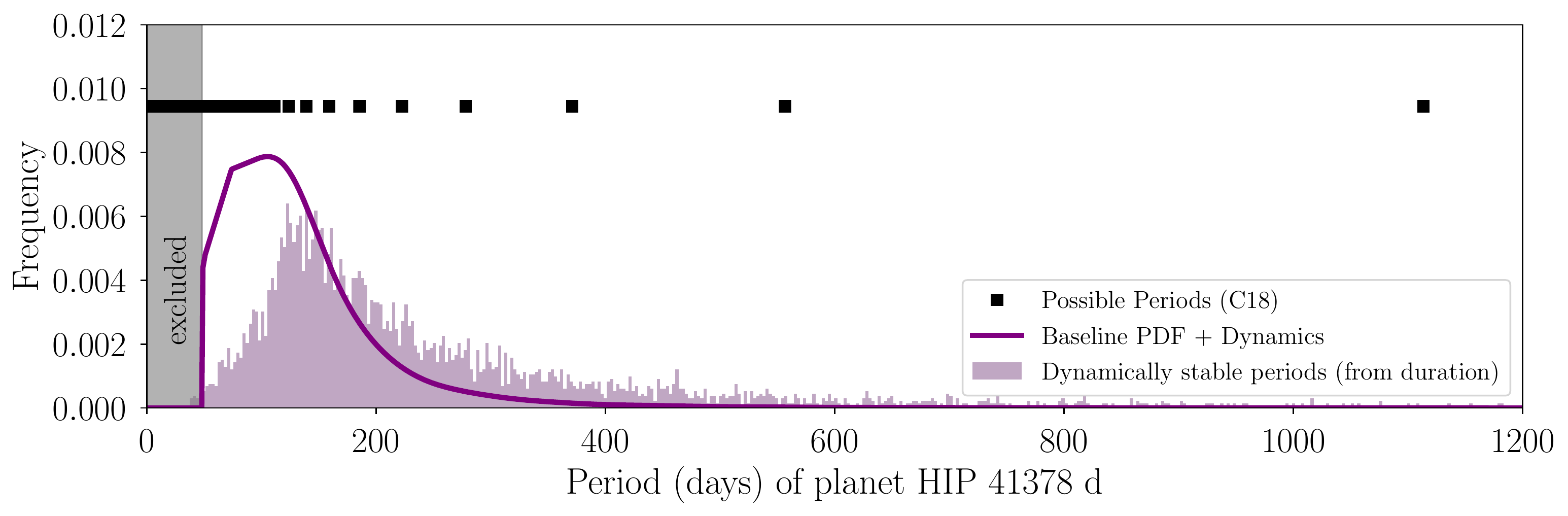} 
      \includegraphics[width=6.0in]{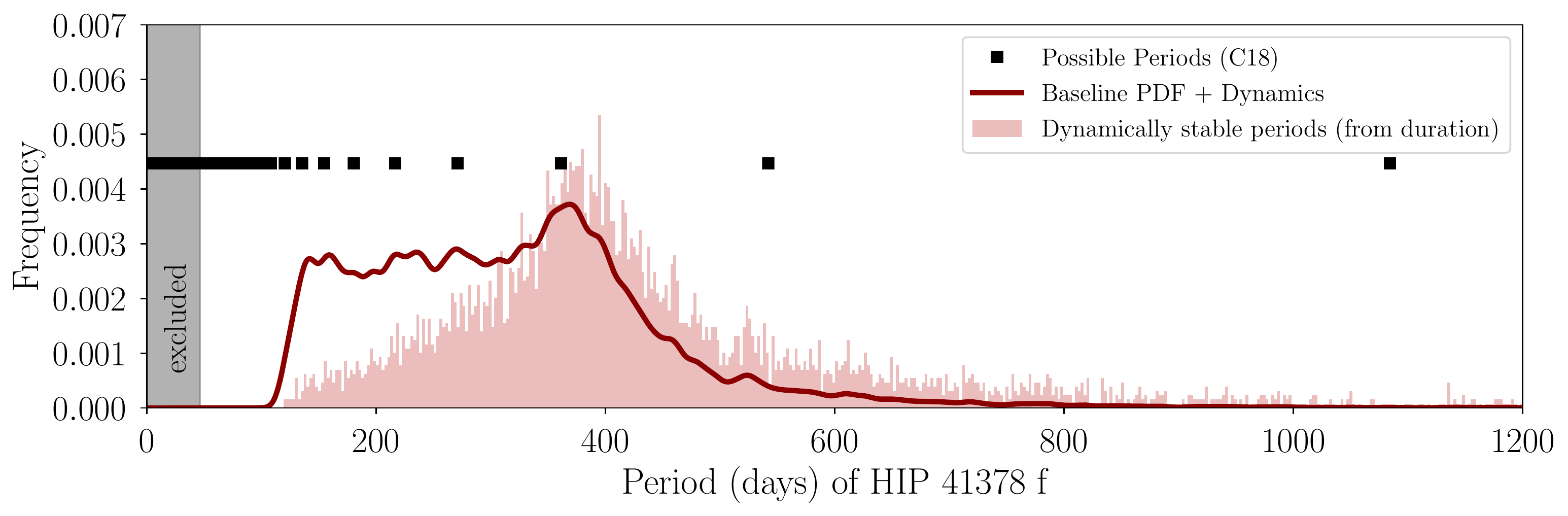} 
   \caption{The derived probability distributions for orbital period for \thisstar\ d (top panel) and \thisstar\ f (bottom panel). The histogram denotes the periods consistent with the measured transit duration, and the height of each bin describes the relative likelihood of each period from dynamical constraints alone. The solid line is the product of the histogram convolved with a Gaussian kernel and the baseline prior (Eq. [\ref{transitlhood}]), which assigns a higher probability to orbital periods with a higher likelihood of transiting during the observed K2 campaigns. At each square point (which correspond to the possible discrete orbital periods), we read off the value of the solid line to get the relative probabilities, which must subsequently be normalized once all possible periods are identified. The results of this analysis at each discrete period are presented in Tables \ref{tab:3} and \ref{tab:5}.}
   \label{fig:both_planets}
\end{figure*}

The orbital periods drawn from this distribution are not necessarily equally physically likely, however. 
To ensure their feasibility, we use two additional criteria based on the dynamical stability of the planetary system as a whole.
V16a used an extensive set of numerical simulations to determine that eccentricities above $e \sim 0.37$ lead to dynamical instability (defined as collisions or ejections within 1 Myr) in the five-planet \thisstar\ system. Additionally, systems are generally expected to be dynamically unstable when their drawn initial conditions are Hill-unstable \citep{fabrycky}.
As such, we exclude from our distribution of dynamically feasible periods any draw which either is Hill-unstable or has planetary eccentricities above 0.37. The result is roughly 5100 orbital periods for each planet that are both consistent with the measured transit duration and adhere to our dynamical stability criterion. A normalized histogram of these orbital periods is shown for each planet in Figure \ref{fig:both_planets}. This histogram represents the probability distribution for the orbital period of each planet, based on only its measured duration and the orbital elements physically likely to cause such a duration. 

We note that for future analysis for other systems, the eccentricity cut we use to exclude dynamically unstable systems will need to be re-derived, as it was derived from numerical simulations for this specific system. The probabilistic exclusion used here will not apply to all systems.

\subsection{Final period constraints for \thisstar\ d and \thisstar\ f}
\label{sec:41}
In this paper, we have constructed various constraints from direct analysis of the light curve, statistical analysis, and dynamical modeling. We list here the constraints that we have placed on the planetary periods:
\begin{itemize}
\item A list of possible periods based on the measured times of transit center for each planet (Eq. \ref{eq:simpleperiods})
\item The statistical baseline prior (Sec. \ref{baseline_out_of_transit}; also see Figure \ref{fig:c18base})
\item Lower limits on the orbital periods from the out-of-transit C5 baseline (Section \ref{baseline_out_of_transit})
\item A distribution of dynamically feasible periods, based on the measured durations and orbital stability
\item Exclusions of particular orbital periods from the combined KELT, HAT, and WASP data (this constraint is available only for \thisstar\ f, which had the deeper transit event)
\end{itemize}
In Figure \ref{fig:both_planets}, we illustrate the final continuous probability distribution with a solid line. This distribution is the normalized product of the baseline transit probability (Eq. \ref{transitlhood}) and the PDF constructed by convolving a Gaussian kernel with the histogram of dynamically feasible periods (generated from Eq. \ref{eq:duration} and described in the previous section). 
Squares denote the possible orbital periods based on Eq. \ref{eq:simpleperiods}, and some of these periods can be excluded using the KELT, HAT, and WASP data. Notably, all of the periods below $\sim$ 100 days can be excluded for \thisstar f.

Using these constraints, we construct individual probability estimates for the possible orbital periods in the following way. First, we exclude all orbital periods generated by Eq. \ref{eq:simpleperiods} that fall below the lower limit derived by Eq. \ref{eq:minper}. Then, for each remaining orbital period, we extract the probability from the interpolated product of the baseline prior and the PDF of dynamically feasible periods (this function is plotted as the solid line in Fig. \ref{fig:both_planets}) at exactly that orbital period. We repeat this for each possible period, and then normalize the total probability for all discrete periods to be equal to one. The resultant periods and their corresponding normalized probabilities are presented in Table \ref{tab:3} and Table \ref{tab:5}.

\subsection{Final period constraints for \thisstar\ e}
\label{sec:planete} 
\begin{figure*} 
   \centering
   \includegraphics[width=6.0in]{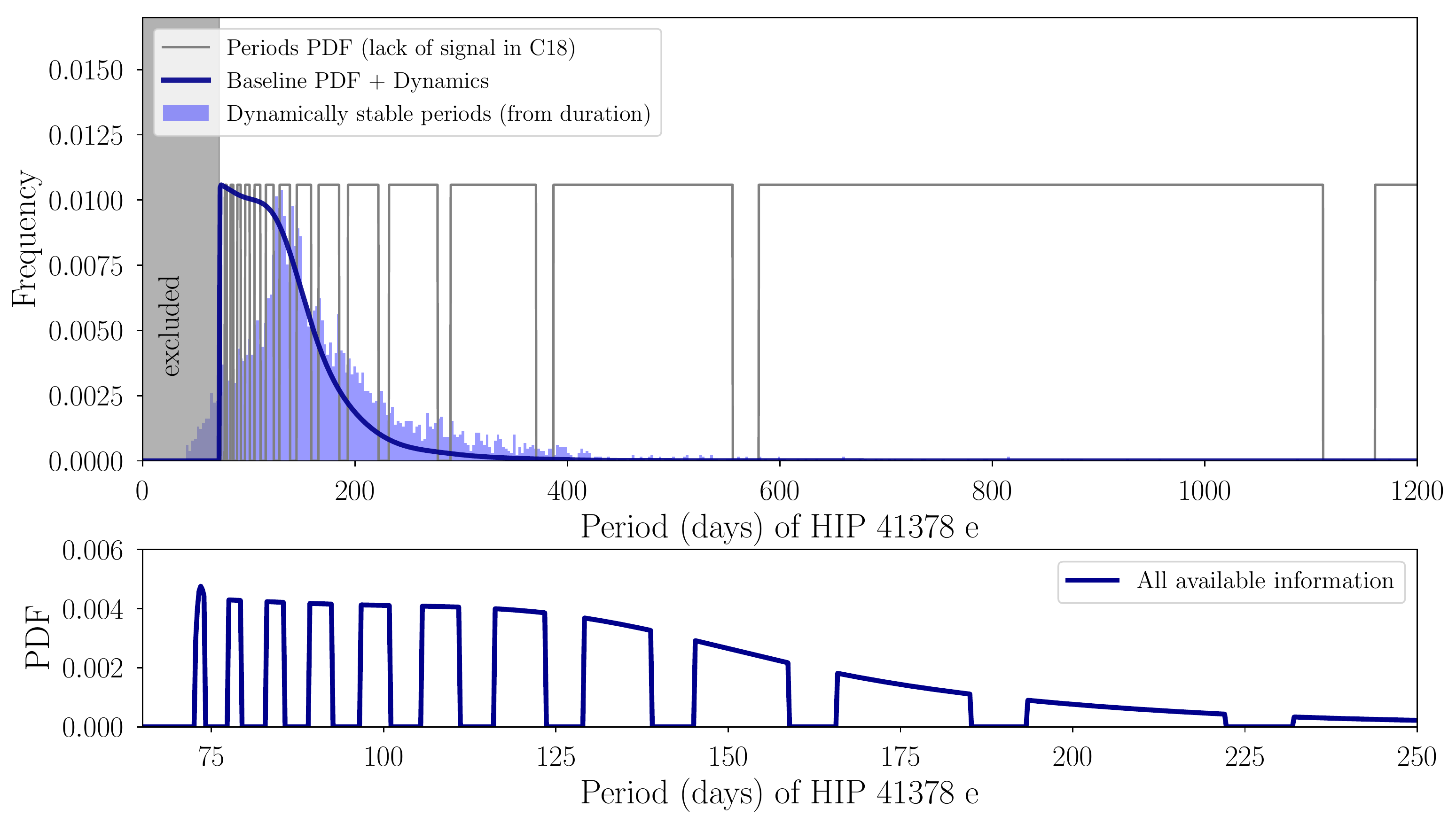} 
   \caption{The derived probability distributions for the orbital period of \thisstar\ e. The histogram (top panel) denotes the periods consistent with the measured transit duration. The solid line is the product of the histogram convolved with a Gaussian kernel and the baseline prior (Eq. [\ref{transitlhood}]), constructed the same way as for \thisstar\ d and \ f in Figure \ref{fig:both_planets}. The grey line describes the relative probability of each orbital period, given that \thisstar\ e did not transit during C18. This line has been normalized so that the maximum value of the Baseline + Dynamics PDF matches its maximum value for illustrative purposes. In the bottom panel, we show the normalized product of the Baseline + Dynamics PDF and the grey curve. }
   \label{fig:eonly}
\end{figure*}

\thisstar\ e transited once during K2 C5, but did not transit during C18 (see Fig. \ref{fig:fulldata}). As such, we do not have discrete guesses for its true orbital period; however, we can exclude any orbital period that would have led to a transit being observable during C18. To construct an additional PDF that represents this scenario, we test each possible orbital period for \thisstar\ e between 72 days (the minimum orbital period permitted by Eq. [\ref{eq:minper}]) and 1200 days. 
Then, we allow $t_{e,18}$ to vary between the times of the first and last data points of C18. 
If Eq. \ref{eq:simpleperiods} is satisfied for some integer $i$ for any value of $t_{e,18}$ on this range, then we consider this particular period ``observable'' in C18, and set the probability that it is the true orbital period of \thisstar\ e to zero. The result of this pruning (normalized so the maximum probability is equal to the maximum probability of the PDF constructed from the results of the baseline PDF and dynamical analysis) is shown in grey in Figure \ref{fig:eonly}.

The final orbital period for \thisstar\ e cannot be directly constrained due to the lack of a transit in the C18 data; the best that can be done without follow-up observations is the probabilistic period estimation presented in the bottom panel of Figure \ref{fig:eonly}.

\begin{deluxetable}{lc}
\tablecaption{Possible orbital periods for \thisstar\ \lowercase{d}}
\tablewidth{0pt}

\tablehead{
  \colhead{Orbital Period (days)} & 
  \colhead{Normalized}   \\
  \colhead{} &
  \colhead{Probability }
}
\startdata
1113.4465 $\pm$ 0.0034 &  $<$ 0.1 \% \\  556.7233 $\pm$ 0.0017 &  $<$ 0.1 \% \\  371.1488 $\pm$ 0.0011 &           0.1 \% \\  278.3616 $\pm$ 0.0009 &           0.5 \% \\  222.6893 $\pm$ 0.0007 &           1.1 \% \\  185.5744 $\pm$ 0.0006 &           2.4 \% \\  159.0638 $\pm$ 0.0005 &           4.1 \% \\  139.1808 $\pm$ 0.0004 &           5.7 \% \\  123.7163 $\pm$ 0.0004 &           6.7 \% \\  111.3447 $\pm$ 0.0003 &           7.1 \% \\  101.2224 $\pm$ 0.0003 &           7.1 \% \\   92.7872 $\pm$ 0.0003 &           7.0 \% \\   85.6497 $\pm$ 0.0003 &           6.9 \% \\   79.5319 $\pm$ 0.0002 &           6.8 \% \\   74.2298 $\pm$ 0.0002 &           6.8 \% \\   69.5904 $\pm$ 0.0002 &           6.3 \% \\   65.4969 $\pm$ 0.0002 &           5.9 \% \\   61.8581 $\pm$ 0.0002 &           5.5 \% \\   58.6024 $\pm$ 0.0002 &           5.1 \% \\   55.6723 $\pm$ 0.0002 &           4.8 \% \\   53.0213 $\pm$ 0.0002 &           4.5 \% \\   50.6112 $\pm$ 0.0002 &           4.2 \% \\   48.4107 $\pm$ 0.0001 &           1.4 \%
\enddata
\tablecomments{Possible orbitals periods and their relative likelihoods, based on the dynamical analysis described in Section\ref{sec:dynamics}. Values may not add up to 100\% due to rounding. Errors on orbital periods were computed with $\sigma = (t_{c,5}^2 + t_{c,18}^2)^{1/2} / n$, when $n$ denotes the number of full cycles between C5 and C18, and $t_{c}$ denotes the uncertainty on center time of transit in each campaign. Errors on the orbital period are lower when a larger number of periods have elapse since the C5 observation.}
\label{tab:3}
\end{deluxetable}

\begin{deluxetable*}{lcc}
\tablecaption{Possible orbital periods for \thisstar\ \MakeLowercase{f}}
\tablewidth{0pt}
\tablehead{
  \colhead{Orbital Period (days)} & 
  \colhead{Normalized }   &
  \colhead{Normalized }   \\
  \colhead{} &
  \colhead{Probability}   &
  \colhead{Probability}   \\
  \colhead{} &
  \colhead{(w/o KELT/HAT/WASP)}   &
  \colhead{(w/ KELT/HAT/WASP)}   }
\startdata
1084.15946 $\pm$ 0.00086 &  $<$ 0.1 \% &  $<$ 0.1 \% \\  542.07973 $\pm$ 0.00043 &           2.2 \% &           3.2 \% \\  361.38649 $\pm$ 0.00029 &          19.9 \% &          29.7 \% \\  271.03986 $\pm$ 0.00022 &          15.7 \% &          23.6 \% \\  216.83189 $\pm$ 0.00017 &          15.2 \% &          22.8 \% \\  180.69324 $\pm$ 0.00014 &          13.4 \% &          20.1 \% \\  154.87992 $\pm$ 0.00012 &          14.8 \% &  $<$ 0.1 \% \\  135.51993 $\pm$ 0.00011 &          13.4 \% &  $<$ 0.1 \% \\  120.46216 $\pm$ 0.00010 &           5.0 \% &  $<$ 0.1 \% \\  108.41595 $\pm$ 0.00009 &           0.4 \% &           0.6 \% \\   98.55995 $\pm$ 0.00008 &  $<$ 0.1 \% &  $<$ 0.1 \% \\   90.34662 $\pm$ 0.00007 &  $<$ 0.1 \% &  $<$ 0.1 \% \\   83.39688 $\pm$ 0.00007 &  $<$ 0.1 \% &  $<$ 0.1 \% \\   77.43996 $\pm$ 0.00006 &  $<$ 0.1 \% &  $<$ 0.1 \% \\   72.27730 $\pm$ 0.00006 &  $<$ 0.1 \% &  $<$ 0.1 \% \\   67.75997 $\pm$ 0.00005 &  $<$ 0.1 \% &  $<$ 0.1 \% \\   63.77409 $\pm$ 0.00005 &  $<$ 0.1 \% &  $<$ 0.1 \% \\   60.23108 $\pm$ 0.00005 &  $<$ 0.1 \% &  $<$ 0.1 \% \\   57.06102 $\pm$ 0.00005 &  $<$ 0.1 \% &  $<$ 0.1 \% \\   54.20797 $\pm$ 0.00004 &  $<$ 0.1 \% &  $<$ 0.1 \% \\   51.62664 $\pm$ 0.00004 &  $<$ 0.1 \% &  $<$ 0.1 \% \\   49.27998 $\pm$ 0.00004 &  $<$ 0.1 \% &  $<$ 0.1 \% \\   47.13737 $\pm$ 0.00004 &  $<$ 0.1 \% &  $<$ 0.1 \% \\
\enddata
\tablecomments{Possible orbitals periods and their likelihoods. The second column comes from only dynamical analysis, and the third column excludes periods that our analysis of the KELT/HAT/WASP data found to be unlikely. Values may not add up to 100\% due to rounding. Errors on orbital periods were computed with $\sigma = (t_{c,5}^2 + t_{c,18}^2)^{1/2} / n$, when $n$ denotes the number of full cycles between C5 and C18, and $t_{c}$ denotes the uncertainty on center time of transit in each campaign.}
\label{tab:5}
\end{deluxetable*}

\label{discussion}

\section{Discussion}
\label{sec:discuss} 

\subsection{Strategies for observational follow-up in the \thisstar\ system}

In this work, we have identified a discrete set of precise possible orbital periods for the long-period transiting planets \thisstar\ d and \thisstar\ f, and have assessed the likelihood that each of these possible orbital periods is indeed the true orbital period. While we have significantly constrained the possible orbital periods of these two planets (we have ruled out about 25\% of the possible periods at confidence \textless1.5\% for planet d and 80\% of possible periods for planet f), our analysis is so far unable to uniquely determine the true orbital periods of these planets. Additional follow-up observations will be necessary to ultimately identify the true orbital periods and enable future studies with facilities like JWST. 

To determine the true orbital periods for \thisstar\ d and f, the strategy is fairly straightforward. The additional transits during C18 and our identification of precise possible orbital periods makes it possible to schedule targeted transit follow-up observations at these most likely periods. The 0.5\% transit depth of \thisstar\ f makes it possible to detect the transit with ground-based telescopes, although the long (19 hour) transit duration will make it impossible to observe the transit from a single observatory. The multi-site Las Cumbres Observatory telescopes, which have demonstrated the ability to produce continuous precise light curves across multiple observing sites around the globe \citep{boyajian2018}, may be well suited to detect the long duration transit of HIP 41378 f. The shallower (800 ppm) transits of HIP 41378 d, however, will likely require space-based resources such as the \Spitzer\ Space Telescope, or potentially the CHEOPS space telescope once it launches in 2019, for confirmation.

Because of HIP 41378 d's shorter orbital period, and the fact that our ground-based data from HAT and KELT were unable to detect or rule out its shallow transits, there are a large number of possible orbital periods, many of which have roughly equal probabilities of being the true orbital period. Observing transits at all of these possible transit times would be an expensive observing program for a precious resource like \Spitzer. However, it should be possible to significantly increase the efficiency of \Spitzer\ follow-up observations for these possible orbital periods because of how many of these periods are related to one another by harmonics. For example, a single \Spitzer\ non-detection of a transit of HIP 41378 d on the observation opportunity on 2019 June 16 (371.149 days after the C18 transit) would rule out four possible orbital periods (371.149, 185.574, 123.716, and 61.858 days). Taking advantage of these harmonic relationships between the possible orbital periods may significantly decrease the amount of observing time needed to identify the true period of HIP 41378 d. 

Determining the orbital period of \thisstar\ e is significantly more difficult than \thisstar\ d and f. Since K2 reobserved \thisstar\ for 51 days and did not detect a second transit of \thisstar\ e, it is likely the orbital period is longer than that of \thisstar\ d, despite their very similar transit durations. The shallow transit depth of only about 0.15\% will likely require long stares with highly precise space-based photometers to redetect. The first opportunity for a re-detection will come fairly soon with the newly commissioned TESS spacecraft. TESS will observe HIP 41378 f in early 2019 (2019 January 7 to 2019 February 2) during Sector 7 of its prime mission\footnote{\url{https://heasarc.gsfc.nasa.gov/cgi-bin/tess/webtess/wtv.py}} and should have sufficient photometric precision to detect the transit of \thisstar\ e. If no transits are detected during the Sector 7 TESS monitoring of \thisstar, TESS may monitor HIP 41378 for a longer period of time in an extended mission, which could provide additional opportunities to detect the transit of this planet. If TESS is unable to re-detect \thisstar\ e, CHEOPS may be able to, if HIP 41378 is added to its monitoring program. The long duration of the transit could make it an efficient CHEOPS target, where only sparse observations are necessary to sample the transit shape. 

\subsection{The uniqueness of \thisstar\ f}

The detection of a second transit of \thisstar\ d and f provides a path towards determining their precise orbital periods and enabling follow-up opportunities for these two long-period gas giant planets. While both planets present intriguing prospects for observations like transmission spectroscopy, \thisstar\ f is a particularly unique target. Depending on its true orbital period, the equilibrium temperature of \thisstar\ f likely ranges between 300 K and 400 K (assuming an albedo similar to Jupiter's), significantly cooler than all other transiting gas giant planets well suited for transmission spectroscopy. We queried the NASA Exoplanet Archive\footnote{\url{http://exoplanetarchive.ipac.caltech.edu/cgi-bin/TblView/nph-tblView?app=ExoTbls&config=planets}} Confirmed Planet Table on 2 September 2018, and identified all transiting planets larger than 0.8 $R_J$ and orbital periods longer than 150 days. Among the nine stars which host planets satisfying these criteria, HIP 41378 is the brightest by a factor of about 15 in H-band. Once a unique transit ephemeris has been determined, the brightness of HIP 41378 should make transmission spectroscopy observations of this long-period temperate gas giant feasible. 

HIP 41378 f will likely remain a uniquely interesting target for transmission spectroscopy into the TESS era as well. We downloaded the predicted TESS planet detection yields from \citet{sullivan} and searched again for planets larger than 0.8 $R_J$ with orbital periods longer than 150 days. Over the course of its two year prime mission, TESS is expected to detect only about three such planets. In the TESS realization from \citet{sullivan}, none of the host stars of these planets are brighter than HIP 41378. It is also likely that any similar long-period planets detected by TESS will have similar orbital period ambiguities to those posed by HIP 41378 \citep[more than 1200 single-transit planets are expected to be found in the full frame images, some of which will have periods longer than 250 days;][]{Villanueva:2018}, so it may be a long time before any long-period TESS discoveries will have precisely determined transit ephemerides to enable follow-up. Now that HIP 41378 f has a straightforward path towards a well determined orbital period, it is likely this planet will be one of the best to study the atmosphere of Jovian planets in temperate, nearly Earth-like irradiation environments.

\subsection{\thisstar\ as a road-map for TESS period determinations}
A additional motivation of the work in this paper is to provide a blueprint for future period-recovery efforts. 
In the era of TESS, many more planets for which the exact orbital period cannot be determined will be discovered. Due to the TESS survey strategy, in some cases, stars will be observed with significant gaps in between periods of observations. For example, according to the Web TESS Viewing Tool\footnote{\url{https://heasarc.gsfc.nasa.gov/cgi-bin/tess/webtess/wtv.py}}, the southern circumpolar star $\delta$ Mensae will be observed by TESS during Sectors 1,5,8,12, and 13 for 28 days each, with gaps of 84 days, 56 days, and 84 days between subsequent periods of observation. Any planet detected by TESS in this region of sky with a period longer than about 28 days could have ambiguous orbital periods due to the observational strategy. The different constraints we used to narrow down the possible orbital periods for the HIP 41378 planets provided in Section \ref{sec:41} can serve as a starting point for future analysis on TESS planets with similar orbital period ambiguities.

Some results, such as the generalized transit probability by transit baseline (Eq. [\ref{transitlhood}]), can be derived merely by substituting in the values for campaign baseline and other easily obtained parameters. Similarly, the expected period distribution can be derived from transit duration (Eq. [\ref{eq:duration}]), as long as sensible priors are applied. For example, in V16a, we pointed out the importance of allowing planets to have non-zero eccentricities when computing period estimates from transit durations: a null eccentricity prior artificially narrows the distribution of possible orbital periods. 
 
On the other hand, some additional constraints may take significantly more work to derive for some systems.
In particular, for multi-planet systems with ambiguous orbital periods, dynamical constraints should be derived uniquely (either using numerical simulations or other dynamical techniques) for each system, and may place tighter limits in some systems than others.

Finally, period estimations can be improved by using additional data. The analysis in this paper also shows the importance of legacy ground-based surveys in the TESS era. Through a combination of existing photometric data from such as HAT and KELT and dynamical analysis, the most likely orbital periods for individual planets can be determined, which allows for an efficient use of limited follow-up telescope resources.

\section{Summary}
\label{sec:conclude}  

In this paper, we have refined the estimates for the orbital periods of \thisstar\ d and \thisstar\ f to provide updated predictions of the transit ephemerides. Although unique orbital periods cannot yet be determined, we have constrained the possibilities and have identified the most likely candidate orbital periods. Additional observations that probe each of the most likely orbital periods will allow a determination of the true orbital period for each planet.  The orbital periods that should be tested are presented in Tables \ref{tab:3} and \ref{tab:5}. One of the primary motivations for this analysis is to recover the transit of \thisstar\ f, a Jupiter-size planet which may be a particularly interesting target for additional in-transit study (such as transit spectroscopy).

The methods developed in this work can be applied to multi-planet systems discovered in the future (e.g., where only single transits are observed). The TESS mission is expected to discover many such systems. In addition, some of the TESS targets are expected to have variable baselines between continuous viewing periods, resulting in a cadence similar to the gaps between campaigns in K2. As a result, true period and ephemeris determinations will be imperative for the subsequent study of many planetary systems discovered by TESS.

\acknowledgments 
During the preparation of this manuscript, we became aware of a parallel paper on the new K2 observations of HIP 41378: Berardo et al. (2018). These manuscripts were prepared independently, and we did not discuss the results with the other team before submission. We thank David Berardo and collaborators for coordinating submission of these papers. 

We thank Tali Khain for her careful reading of the manuscript and useful suggestions. 

J.C.B is supported by the NSF Graduate Research Fellowship, grant No. DGE 1256260. This work was performed in part under contract with the California Institute of Technology/Jet Propulsion Laboratory funded by NASA through the Sagan Fellowship Program executed by the NASA Exoplanet Science Institute.
A.C. and D.H. acknowledge support by the National Aeronautics and Space Administration under Grants NNX17AF76G and 80NSSC18K0362 issued through the K2 Guest Observer Program.

This research has made use of NASA's Astrophysics Data System and the NASA Exoplanet Archive, which is operated by the California Institute of Technology, under contract with the National Aeronautics and Space Administration under the Exoplanet Exploration Program. This work used the Extreme Science and Engineering Discovery Environment (XSEDE), which is supported by National Science Foundation grant number ACI-1053575. This research was done using resources provided by the Open Science Grid, which is supported by the National Science Foundation and the U.S. Department of Energy's Office of Science.

This paper includes data collected by the \Kepler\ mission. Funding for the \Kepler\ mission is provided by the NASA Science Mission directorate. Some of the data presented in this paper were obtained from the Mikulski Archive for Space Telescopes (MAST). STScI is operated by the Association of Universities for Research in Astronomy, Inc., under NASA contract NAS5--26555. Support for MAST for non--HST data is provided by the NASA Office of Space Science via grant NNX13AC07G and by other grants and contracts.

This work has made use of data from the European Space Agency (ESA) mission
{\it Gaia} (\url{https://www.cosmos.esa.int/gaia}), processed by the {\it Gaia}
Data Processing and Analysis Consortium (DPAC,
\url{https://www.cosmos.esa.int/web/gaia/dpac/consortium}). Funding for the DPAC
has been provided by national institutions, in particular the institutions
participating in the {\it Gaia} Multilateral Agreement.

Facilities: \facility{Kepler/K2, KELT, HATNet, WASP}
 
Software: pandas \citep{mckinney-proc-scipy-2010}, IPython \citep{PER-GRA:2007}, matplotlib \citep{Hunter:2007}, scipy \citep{scipy}, numpy \citep{oliphant-2006-guide}, Jupyter \citep{Kluyver:2016aa}, FITSH \citep{2011ascl.soft11014P, 2012MNRAS.421.1825P}, Kadenza \citep{kadenza}

\bibliographystyle{apj}



\end{document}